\documentclass[prb,preprintnumbers,superscriptaddress,aps,showpacs]{revtex4}

\begin{document}
\title{Non-linear $\sigma $-model for odd triplet superconductivity in
superconductor/ferromagnet structures }
\author{J. E. Bunder}
\affiliation{Physics Division, National Center for Theoretical Sciences, Hsinchu 300, Taiwan}
\author{K. B. Efetov}
\affiliation{Theoretische Physik III, Ruhr-Universit\"at-Bochum,
D-44780 Bochum, Germany} \affiliation{L. D. Landau Institute for
Theoretical Physics, 117940 Moscow, Russia}

%\author{J. E. Bunder${}^{1}$ and K.B. Efetov$^{2,3}$}
%\address{${}^1$ Physics Division, National Center for %Theoretical Sciences, 101
%Section 2 Kuang Fu Road, Hsinchu 300, Taiwan \\
%${}^2$ Theoretische Physik III, \\
%Ruhr-Universit\"at Bochum, 44780 Bochum, Germany\\
%${}^3$ L. D. Landau Institute for Theoretical Physics, 117940 %Moscow, Russia}

\date{\today}

\begin{abstract}
We consider some properties of odd frequency triplet superconducting
condensates. In order to describe fluctuations we construct a supermatrix $%
\sigma $-model for the superconductor/ferromagnet or
superconductor/normal-metal structures. We show that an odd frequency
triplet superconductor, when in isolation or coupled to a normal metal,
generally displays behaviour comparable to a superconductor with the usual
singlet pairings. However, for spin dependent systems such as the
superconductor/ferromagnet the two types of superconductor have quite
different behaviour. We discuss this difference by considering
transformations under which the $\sigma $-model is invariant. Finally, we
calculate the low energy density of states in a ferromagnet coupled to a
singlet superconductor. If odd frequency triplet components are induced in
the ferromagnet the density of states will have a micro-gap similar to the
micro-gap found in normal metals coupled to a superconductor.
\end{abstract}

\maketitle

%\address{${}^1$ Theoretische Physik III, \\
%Ruhr-Universit\"at Bochum, 44780 Bochum, Germany\\
%${}^2$ L. D. Landau Institute for Theoretical Physics, 117940 Moscow, Russia}

%\draft

%74.50.+r Proximity effects, weak links, tunneling phenomena,
% and Josephson effects
%74.20.Rp Pairing symmetries (other than s-wave)
%74.25.Fy   Transport properties (electric and thermal conductivity,
%thermoelectric effects, etc.)
%73.23.-b     Electronic transport in mesoscopic systems

\section{Introduction}

The Pauli principle imposes important restrictions on the symmetry of the
superconducting condensate in superconductors. The most common condensate is
a singlet where the Cooper pairs have antiparallel spins ($s$- or $d$-wave).
In this case, the wave function describing Cooper pairs is assumed to be
invariant under the exchange of electron coordinates. Another possibility is
a triplet pairing with the total spin of the pair equal to unity. In this
case the wave function of the pair is assumed to change sign if the
electrons exchange coordinates. The most famous example of the triplet
pairing ($p$-wave) is superfluid ${\rm He}^{3}$ \cite{helium} but triplet
superconductivity has been recently discovered.\cite{triplet1,triplet2}

However, a characterization of the superconductor in terms of space
symmetries of the wave function of Cooper pairs is somewhat oversimplified.
The full information about the superconducting condensate is in fact given
by an %quasiclassical 
anomalous Green function (Gorkov function) $F(\epsilon)$. 
%$f_{\epsilon }$. 
This function depends not only on the coordinates of the Cooper pair but
also on the frequency $\epsilon$. The previous discussion about the
properties of the wave function of the Cooper pairs corresponds to the case
when the condensate function $F(\epsilon)$\ is an even function of the
frequency $\epsilon $ although nothing forbids the function $F(\epsilon)$
from being an odd function of $\epsilon $. If this alternative possibility
were realized one would have a situation where the condensate function $%
F(\epsilon)$ is invariant under the permutation of electrons with triplet
pairing but would change sign in the singlet case. So, odd condensate
functions of frequencies allow, at least theoretically, $p$-wave singlets
and $s$- and $d$-wave triplets.

In this paper we shall discuss some aspects of triplet Cooper pairings
which are odd in frequency and even in momentum. A superconductor with an
odd frequency triplet condensate was introduced by Berezinskii~\cite
{berezinskii} as a possible candidate for a phase of ${\rm He}^{3}$, though
this was later found to not be the case. One may also consider other
symmetry variations. For example, in Ref. \onlinecite{balatsky} an odd
singlet superconductor (one which is odd in both frequency and momentum) was
discussed. Unfortunately, the authors of Refs. %
\onlinecite{berezinskii,balatsky} did not find a microscopic model that
would lead to the odd frequency condensate.

Recently, it was found that the odd triplet condensate can be induced in a
superconductor/ferromagnet structure provided the magnetization in the
ferromagnet is inhomogeneous.~\cite{bergeret} In this situation one does not
need a special kind of an electron-electron interaction. It is sufficient
that the ferromagnet is coupled to a standard singlet superconductor. This
shows that, independent of whether the odd superconductivity can be obtained
as the ground state of a microscopic model or not, a detailed study of its
properties based just on the symmetry of the condensate may be of interest
because it can be realized at least as a proximity effect.

In this paper we compare properties of the odd triplet superconductivity
with those of the conventional singlet. We first consider a superconductor
with odd frequency triplet pairings ($S_{t}$). We construct the Gorkov
Green functions and write them in terms of an integral over supervectors,
which allows us to obtain a supermatrix $\sigma $-model. It turns out that
the form of the Green functions closely resembles those of a standard
singlet superconductor ($S_{s}$). In fact, one can show that in many cases an 
$S_{t}$ will have very similar properties to an $S_{s}$. Differences appear
when one considers spin dependent structures such as a superconductor
coupled to a ferromagnet ($S_{s}/F$ or $S_{t}/F$). These two types of
superconductors have different symmetries of the order parameter which leads
to differences in the Josephson effect. A qualitative discussion about the
proximity effect in $S_{t}/F$ structures may be made from considering
transformations under which the $\sigma $-model is invariant. From these
transformations one can determine which types of Cooper pairs are induced in
the ferromagnet and whether the penetration is long-ranged or short-ranged.
Generally, it is simpler to just solve the saddle point equation, but if the
ferromagnet has a complicated inhomogeneous structure consideration of the
transformational invariances may be useful.

It is well known that the density of states of an $S_{s}$ in isolation (and
also an $S_{t}$) has an energy gap equal to the value of the order
parameter. A normal metal has no energy gap. However, in an $S_{s}/N$
structure it has been shown that the density of states in the normal metal
decreases quadratically at low energies and vanishes completely at zero
energy.~\cite{frahm,altland} This region of vanishing density of states is
called the `micro-gap' and is a consequence of long-ranged Cooper pairs
being induced in the normal metal. For most $S_{s}/F$ structures no
long-ranged Cooper pairs are induced in the ferromagnet and so there is no
micro-gap comparable to the micro-gap found in $S_{s}/N$ structures.
However, as a result of some inhomogeneities in the ferromagnets (domain
walls can be an example), an odd triplet condensate may be induced. \cite
{bergeret} We consider such a $S_{s}/F$ structure and calculate the low
energy C-mode fluctuations about the saddle point solution. From this we can
calculate the density of states. We find that the density of states in the
ferromagnet has a micro-gap similar to the micro-gap found in $S_{s}/N$
structures. As concerns a $S_{t}/F$ structure, the superconducting
condensate will always penetrate the ferromagnet, even if the ferromagnet is
homogeneous and this penetration will always be long-ranged. As a result, we
would expect the ferromagnet in an $S_{t}/F$ structure to always exhibit a
micro-gap.

\section{Gorkov Green functions}

In this section we construct the Green functions for an odd frequency
triplet condensate and compare it to the Green functions of an even
frequency singlet. We begin with the general form of the superconductor
Hamiltonian 
\begin{equation}
H=\int dr \left[\psi _{\alpha }^{\dag }(r){\cal H} (r)\psi _{\alpha }(r)
+\psi _{\alpha }^{\dag }(r)V_{\alpha \beta }(r)\psi _{\beta }(r)+{\textstyle{%
\frac{1}{2}}}\int dr^{\prime }\psi _{\delta }^{\dag }(r)\psi _{\gamma
}^{\dag }(r^{\prime })U_{\delta \gamma \alpha \beta }(r,r^{\prime })\psi
_{\alpha }(r^{\prime })\psi _{\beta }(r)\right]
\end{equation}
where ${\cal H}$ is the one-particle Hamiltonian, $V_{\alpha \beta }$ is the
exchange field which may have some spatial dependence, $U_{\delta \gamma
\alpha \beta }$ is the two-particle potential and $\psi _{\xi }$ and $\psi
_{\xi }^{\dag }$ are fermionic destruction and annihilation operators. This
form of the Hamiltonian is completely general with regards to spin, time and
position symmetries. Since $H$ must be Hermitian $V=V^{\dag }$ and $%
U=U^{\dag }$. In coupled systems such as $S/F$ and $S/N$ the superconductor
is defined to lie along the negative $x$-axis and the ferromagnet or
normal-metal lies along the positive $x$-axis. The two-particle potential $U$
is just defined in the superconductor so vanishes in $F$ and $N$. The
exchange field vanishes in $S$ and $N$. From the above Hamiltonian and using
the conventional mean field approximation we can construct the Green
functions that have both normal and anomalous components. For more details
see for example Ref. \onlinecite{abrikosov}. The dynamic equations for $\psi
_{\xi }$ and $\psi _{\xi }^{\dag }$ are obtained from the identity $\frac{%
\partial \Phi }{\partial t}=i[H,\Phi ]$. The Green function $G_{\alpha
\beta }\left( X,X^{\prime }\right) $ and the anomalous Green function $%
F_{\alpha \beta }\left( X,X^{\prime }\right) $ are defined by 
\begin{eqnarray}
G_{\alpha \beta }(X,X^{\prime }) &=&-\langle T\Psi _{\alpha }(X)\Psi _{\beta
}^{\dag }(X^{\prime })\rangle ,  \nonumber \\
F_{\alpha \beta }(r,r^{\prime }) &=&-\langle T\Psi _{\alpha }(X)\Psi _{\beta
}(X^{\prime })\rangle ,  \nonumber \\
F_{\alpha \beta }^{\dag }(r,r^{\prime }) &=&-\langle T\Psi _{\alpha }^{\dag
}(X^{\prime })\Psi _{\beta }^{\dag }(X)\rangle ,
\end{eqnarray}
where $T$ is the time-ordering operator, $X=\left( r,t\right) $ and $\Psi $
and $\Psi ^{+}$ are the operators in the Heisenberg representation. Usually
one complements the Green function equations with the self-consistency
equation 
\begin{equation}
\Delta _{\xi \alpha }(X,X^{\prime })=U_{\xi \alpha \gamma \beta
}(X,X^{\prime })\langle \Psi _{\gamma }(X^{\prime })\Psi _{\beta }(X)\rangle.
\label{e1}
\end{equation}
Because of the Pauli exclusion principle the anomalous Green functions $F$
must be anti-symmetric under simultaneous position-time and spin exchange.
It follows from Eq. (\ref{e1}) that the order parameter $\Delta $ has the
same symmetry as the anomalous Green function $F$. In the singlet state $%
S_{s}$ the order parameter is even in time and position exchange. In the
triplet state $S_{t}$ considered here the order parameter is odd in time
exchange but even in position exchange. However, we emphasize that we cannot
and do not try to present a microscopic model that would determine the odd
triplet superconducting order parameter $\Delta $ but write it purely
phenomenologically. Note that the odd triplet condensate function $F$ can
exist due to the proximity effect in $S/F$ structures.\cite{bergeret}

We use the dynamic equations of $\Psi _{\xi }$ and $\Psi _{\xi }^{\dag }$
and the definitions for the Green functions to write dynamic equations for
the Green functions. In order to simplify the symmetry considerations we
chose the order parameter to be $\Delta (X,X^{\prime })=\Delta
(r,t,t^{\prime })\delta (r-r^{\prime })$.\footnote{%
In the case of an $S_{s}$ we can take $U_{\delta \gamma \alpha \beta
}(r,r^{\prime \prime })\propto \delta (r-r^{\prime \prime })$, though not
for an $S_{t}$ as this will destroy the odd symmetry. The symmetry in $r$
and $r^{\prime }$ implies we are considering an $s$-wave} For the
conventional singlet superconductivity the function $\Delta \left(
r,t,t^{\prime }\right) $ is invariant under the exchange of $t$ and $%
t^{\prime}$ whereas in the triplet case considered here it changes sign.
After taking a Fourier transform the advanced and retarded Gorkov Green
functions represented in particle-hole space are 
\begin{eqnarray}
\left( 
\begin{array}{cc}
\epsilon\pm i\delta/2 -{\cal H}-V & \Delta (x,\epsilon ) \\ 
(-1)^{S+1} \Delta ^{\ast }(x,-\epsilon ) & -\epsilon\mp i\delta/2 -{\cal H}%
-V^{\ast }
\end{array}
\right) {\cal G}^{R,A}(x,x^{\prime },\epsilon ) &=&\delta (x-x^{\prime }), 
\nonumber \\
{\cal G} &=&\left( 
\begin{array}{cc}
G & F \\ 
F^{\dag } & G^{\dag }
\end{array}
\right) ,  \label{eq:gorkov}
\end{eqnarray}
where $S$ is the total spin of the Cooper pair and $\delta$ is a small
positive real number, the sign in front of which determines the advanced or
retarded nature of the Green function.  
We see that the difference between the equations for the conventional
singlet and odd triplet superconductivities is minimal. Note that the spin
dependence is hidden inside $G$, $F$, $\Delta $ and $V$.

If the spin is represented by the Pauli matrices $\sigma $ we can expand the
order parameter as $\Delta =\sum_{i=0}^{3}\Delta _{i}\sigma _{i}$ and we may
write each component in terms of a phase, $\Delta _{i}=|\Delta
_{i}|e^{i\theta _{i}}$. We represent the triplet components of $\Delta $ by $%
\sigma _{0}$, $\sigma _{1}$ and $\sigma _{3}$ and the singlet ones by $%
\sigma _{2}$. With this choice we satisfy the symmetry relations $\Delta
=-\Delta ^{T}$ for the conventional singlet superconductivity and $\ \Delta
=\Delta ^{T}$ for the odd triplet. For conventional even frequency
superconductors the order parameter is often assumed to be energetically
independent, however, in the case of an $S_{t}$ the order parameter must be
odd in energy so we choose the simplest possibility $\Delta (x,\epsilon )=\,%
{\rm sgn}(\epsilon )\Delta (x)$.

In order to study mesoscopic fluctuations we use the supersymmetry method. 
\cite{efetov} Within this technique one can write the solution of Eq. (\ref
{eq:gorkov}) in terms of a functional integral over supervectors $\psi $~
\cite{kravtsov,altland,taras,efetov} 
\begin{eqnarray}
{\cal G}^{R,A}(x,x^{\prime },\epsilon ) &=&i\int \psi _{\alpha
}^{2,1}(x)\otimes \bar{\psi}_{\alpha }^{2,1}(x^{\prime })\exp [-{\cal L}%
_{s,t}]{\cal D}\psi   \nonumber \\
{\cal L}_{s} &=&i\int \bar{\psi}(y)\left( 
\begin{array}{cc}
\epsilon -i\delta \Lambda /2-{\cal H}-V & \Delta (y) \\ 
-\Delta ^{\ast }(y) & -\epsilon +i\delta \Lambda /2-{\cal H}-V^{\ast }
\end{array}
\right) \psi (y)dy  \nonumber \\
{\cal L}_{t} &=&i\int \bar{\psi}(y)\left( 
\begin{array}{cc}
\epsilon -i\delta \Lambda /2-{\cal H}-V & \,{\rm sgn}(\epsilon )\Delta (y)
\\ 
-\,{\rm sgn}(\epsilon )\Delta ^{\ast }(y) & -\epsilon +i\delta \Lambda /2-%
{\cal H}-V^{\ast }
\end{array}
\right) \psi (y)dy  \label{eq:susy}
\end{eqnarray}
where ${\cal L}_{s,t}$ is the action for the singlet and the odd triplet
superconductivity respectively and all other terms have the standard
definitions. If we perform the gauge transformation $\psi \rightarrow \psi
e^{i\frac{\pi }{4}[\,{\rm sgn}(\epsilon )-1]\tau _{3}}$ and $\bar{\psi}%
\rightarrow \bar{\psi}e^{-i\frac{\pi }{4}[\,{\rm sgn}(\epsilon )-1]\tau _{3}}
$ where $\tau $ represents Pauli matrices of the particle-hole space we find
that, if we ignore the spin dependence, the triplet action is no different
from the singlet action but the coefficient of the exponential becomes $%
[\psi _{\alpha }^{2,1}(x)\otimes \bar{\psi}_{\alpha }^{2,1}(x^{\prime
})]_{mn}\rightarrow \lbrack \psi _{\alpha }^{2,1}(x)\otimes \bar{\psi}%
_{\alpha }^{2,1}(x^{\prime })]_{mn}[\,{\rm sgn}(\epsilon )]^{m-n}$ where $m$
and $n$ represent components of the particle-hole space. Thus, if spin is
not important the normal odd triplet Green functions $G$ are identical to
the normal singlet Green functions but the anomalous triplet Green
functions $F$ differ from that of the singlet by a factor of $\,{\rm sgn}%
(\epsilon )$, i.e., the singlet's anomalous Green functions are even in $%
\epsilon $ but the triplet's are odd, as expected from the initial symmetry
requirements. As the normal Green functions determine the density of
states the bulk singlet and the bulk triplet have the same density of
states. Also, a $S_{t}/N$ structure should be similar to a $S_{s}/N$
structure since in these cases spin is not important.

\section{Transformational invariances of the $\protect\sigma$-model}

From Eq. (\ref{eq:susy}) the construction of a $\sigma $-model is fairly
straight forward. Using the standard method of derivation developed for the
singlet superconductor the $\sigma $-model action may be shown to be~\cite
{altland,kravtsov} 
\begin{equation}
S={\textstyle{\frac{\pi \nu }{16}}}\,{\rm str}\,\int [D(\partial
Q)^{2}+4iQ(\epsilon \tau _{3}-i\delta \Lambda \tau _{3}/2-\tilde{\Delta}-\,%
{\rm Re}\,V-i\tau _{3}\rho _{3}\,{\rm Im}\,V)].  \label{eq:action}
\end{equation}
where $\rho _{3}$ is the third Pauli matrix in the time-reversal space, $Q$
is a $32\times 32$ supermatrix, $\nu$ is the bulk normal-metal density of
states per spin and 
\begin{equation}
\tilde{\Delta}=i\tau _{2}\rho _{3}[\sigma _{0}|\Delta _{0}|\exp (-i\theta
_{0}\tau _{3}\rho _{3})+\sigma _{1}|\Delta _{1}|\exp (-i\theta _{1}\tau
_{3}\rho _{3})+\sigma _{3}|\Delta _{3}|\exp (-i\theta _{3}\tau _{3}\rho
_{3})]-\sigma _{2}\tau _{1}\rho _{3}|\Delta _{2}|\exp (-i\theta _{2}\tau
_{3}\rho _{3}).
\end{equation}
The $Q$-matrices in Eq. (\ref{eq:action}) must satisfy as usual the charge
conjugation symmetry and integrals with the action $S$ must converge. In
addition, one can find several transformations under which $Q$ is invariant
in the bulk superconductor (when $V=0$).~\cite{frahm} We define $A$ to be
invariant under the transformation $C$ if $A=CA^TC^T$. Table \ref{symmtable}
defines five transformations and the terms with which they are invariant.
All the terms in the action of a triplet superconductor are invariant under
the $C_{4}$ transform while the singlet superconductor action is invariant
under the other four transforms. This appears to disagree with what was
found in Ref. \onlinecite{frahm} where it was claimed that the singlet was
invariant under the $C_{4}$ transform. The difference is due to the spin
dependence of our $\sigma $-model. In general the ferromagnetic exchange
field is of the form $V=h_{0}\sigma _{0}+h_{1}\sigma _{1}+h_{2}\sigma
_{2}+h_{3}\sigma _{3}$ (all the $h_{i}$ must be real since $V=V^{\dag }$).
In the ferromagnet $Q$ is not required to be invariant under any of the
transforms in table \ref{symmtable} but they can help in determining the
form of $Q$ in the ferromagnet.

\begin{table}[tbp]
\caption{Transformational invariances of the $\protect\sigma$-model action.
The matrix $A$ is invariant under the transform $C$ if $A=CA^TC^T$. The
singlet superconductor action is invariant under the $C_0$, $C_1$, $C_2$ and 
$C_3$ transforms whereas the triplet superconductor action is only invariant
under the $C_4$ transform. In addition both types of superconductor actions
must have charge conjugation and convergence symmetry.}
\label{symmtable}
\begin{tabular}{c|c}
\hline
transform & invariance \\ \hline
$C_0=i\tau_1$ & $\sigma_2,\, \tau_1\sigma_2,\, \tau_2\sigma_2,\,
\tau_3\sigma_{0,1,3}$ \\ 
$C_1=\tau_2\sigma_1$ & $\sigma_3,\, \tau_1\sigma_{0,1,2},\,
\tau_2\sigma_{0,1,2},\, \tau_3\sigma_{0,1,2}$ \\ 
$C_2=\tau_1\sigma_2$ & $\sigma_{1,2,3},\, \tau_1\sigma_{1,2,3}, \,
\tau_2\sigma_{1,2,3},\, \tau_3$ \\ 
$C_3=\tau_2\sigma_3$ & $\sigma_1,\, \tau_1\sigma_{0,2,3},\,
\tau_2\sigma_{0,2,3},\, \tau_3\sigma_{0,2,3}$ \\ 
$C_4=\tau_2$ & $\sigma_2,\, \tau_1\sigma_{0,1,3},\, \tau_2\sigma_{0,1,3},\,
\tau_3\sigma_{0,1,3}$\\
\hline
\end{tabular}
\end{table}

As an example on how to use the transformational invariances we consider an $%
S_{t}/F$ structure with different exchange fields. The saddle-point equation
of a superconductor $\sigma $-model is also known as the Usadel equation.
The quasiclassical Green function which satisfies the Usadel equation is
the saddle point solution of the $\sigma $-model and is represented by 
\begin{equation}
g_{0}=\left( 
\begin{array}{cc}
g & f \\ 
f^{\dag } & g^{\dag }
\end{array}
\right)
\end{equation}
in the particle-hole space with the constraint $g_{0}^{2}=1$. If we assume
the temperature is just below the superconducting transition temperature or
the tunneling resistivity is very large the Green function in the
ferromagnet is $g^{A,R}=-g^{A,R\dag }\sim \mp 1$. In this case the Usadel
equation may be linearized and the retarded anomalous triplet Green
function can be shown to satisfy 
\begin{equation}
iD\partial _{x}^{2}f-2\epsilon f+Vf-fV^{\ast }=0  \label{eq:usadel}
\end{equation}
in the ferromagnet (having dropped the superscript). This is the same as the
linearised equation in the ferromagnetic region of an $S_{s}/F$ structure
but, due to the boundary conditions, the spin structure of $f$ must be
different.~\cite{bergeret,buzdin} The boundary conditions at the interface
are 
\begin{eqnarray}
\partial _{x}f(0^{+}) &=&[\rho (+)/R_{b}]f(0^{-}),\qquad T\ll 1,  \nonumber
\\
f(0^{+}) &=&f(0^{-}),\qquad \qquad \qquad T\sim 1,
\end{eqnarray}
where `$-$' is the superconducting side of the interface and `$+$' is the
ferromagnetic side, $T$ is the transparency of the interface, $\rho (\pm )$
is the resistivity and $R_{b}$ is the tunneling resistivity. As $%
x\rightarrow -\infty $ the Green function must approach the bulk
superconductor solution and as $x\rightarrow \infty $ it must approach the
bulk ferromagnet solution. Assuming that the proximity effect on the
superconductor is small the well known bulk solution may be taken in the
entire superconducting region $x<0$ where $V=0$ so $f(x<0)=\,{\rm sgn}%
(\epsilon )\Delta /\sqrt{\epsilon ^{2}-|\Delta |^{2}}$. This is the same
solution as for a bulk $S_{s}$ but with the extra term $\,{\rm sgn}(\epsilon
)$ which gives the required odd energy dependence and $\Delta $ has a
different spin dependence. In the ferromagnetic region $x>0$ the anomalous
Green function is of the form $f(x>0)=\sum_{i=0}^{3}f_{i}(x)\sigma _{i}$
(assuming we have both triplet and singlet components). The boundary
condition at $x\rightarrow \infty $ is that all the $f_{i}$ must vanish.

If the magnetisation is of the form $V=h\sigma_j$, $j=1,2,3$ then the
solution of the linearised Usadel equation is that each $f_i$ will
exponentially decay. Two components will decay at a rate independent of the
exchange field, $\kappa_{\epsilon}$ and the other two will dacay at the rate 
$\kappa=\sqrt{\kappa_{\epsilon}^2\pm\kappa_h^2}$ where $\kappa_{%
\epsilon}^2=-2i\epsilon/D$ and $\kappa_h^2=-2ih/D$. For example, if $%
V=h\sigma_3$ the $\sigma_3$ and $\sigma_0$ components of the anomalous
Green function decay at the rate $\kappa_{\epsilon}$ while the $\sigma_1$
and $\sigma_2$ components decay at the rate $\kappa$. When $h$ is large, as
it generally is in such structures, the $\sigma_{0,3}$ components are
long-ranged while the other two are short ranged. The boundary conditions at
the interface require that the $\sigma_2$ component vanishes at the
interface. Inducing long-ranged triplet components $\sigma_{0,3}$ in the
ferromagnet of a $S_t/F$ structure with exchange field $h\sigma_3$ should
not be surprising. However, if $V=h\sigma_2$ we find that the $\sigma_0$ and 
$\sigma_2$ components decay rapidly at the rate $\kappa$ and the $\sigma_3$
and $\sigma_1$ components decay slowly at the rate $\kappa_{\epsilon}$. The
boundary conditions at the interface will make the $\sigma_2$ component
vanish at the interface. In contrast, boundary conditions in an $S_s/F$
structure with a homogeneous ferromagnet potential only allow the $\sigma_2$
anomalous component in the ferromagnet which always decays rapidly at the
rate $\kappa$.

It is easiest to find which $f_{i}$ will decay at which rate from the Usadel
equation. However, one can also determine this from the transformational
invariances of the $\sigma $-model. In more complicated cases it is easier
to determine which components will be significant from the invariances
rather than the Usadel equation, although the Usadel equation is required
for quantitative results. One can show that the anomalous components which
are invariant under the same transform as the $\sigma $-model will have the $%
h$ dependent decay $\kappa $. For example, if we have $S_{t}/F$ with $%
V=h\sigma _{3}$ the transforms under which the associated $\sigma $-model is
invariant are $C_{1}$ and $C_{2}$. These invariances are shared by $\tau
_{1,2}\sigma _{1,2}$ so one may conclude that the $\sigma _{1}$ and $\sigma
_{2}$ components of the anomalous Green functions decay at the $h$
dependent rate $\kappa $. The other two anomalous components, $\sigma _{0}$
and $\sigma _{3}$ are not invariant under the $C_{1}$ and $C_{2}$
transformations so decay at the rate $\kappa _{\epsilon }$ which is
independent of $h$. Similarly, if $V=h\sigma _{2}$ the action is invariant
under the $C_{1}$ and $C_{3}$ transforms, as are the terms $\tau
_{1,2}\sigma _{0,2}$. Therefore the $\sigma _{0,2}$ components of the
anomalous Green functions are short-ranged, decaying at the rate $\kappa $%
, while the other two components $\sigma _{1,3}$ are long-ranged, decaying
at the rate $\kappa _{\epsilon }$.

One case of particular interest is when a superconductor is coupled to an
inhomogeneous ferromagnet. It has been shown that at an $S_s/F$ interface it
is possible to induce both a singlet and an odd frequency triplet component
in the ferromagnet if, for example, $V=h(\sigma_3\cos\alpha+\sigma_2\sin%
\alpha)$.~\cite{bergeret} Here $\alpha=Ax$ for some constant $A$ when $0<x<w 
$ and $\alpha=Aw$ when $x>w$ where $w$ is some positive constant. We shall
briefly describe how the anomalous components induced in the ferromagnet may
be determined from the transformational invariances of the action. At the
interface the ferromagnet potential introduces the term $\tau_0\sigma_3$
into the action so at this point the action is invariant under the $C_1$ and 
$C_2$ transforms. As $x$ increases a $\tau_3\sigma_2$ component appears in
the action. Now the action is invariant only under the $C_1$ transform.
Invariance under the $C_1$ and $C_2$ transforms at the interface implies
short-ranged (decay is $h$ dependent) anomalous components $\sigma_{1,2}$
and long-ranged (decay is $h$ independent) components $\sigma_{0,3}$.
However, as $x$ increases we lose the invariance under the $C_2$ transform.
When $C_1$ is the only transformational invariance the short-ranged
components are $\sigma_{0,1,2}$ and only $\sigma_3$ is long-ranged. However,
the boundary conditions cause the coefficient of the $\sigma_3$ component to
vanish. We may conclude that, if the total rotation $Aw$ is small the
solution within the domain wall will be approximately similar to the
solution at the $S_s/F$ interface. Thus we would expect the $\sigma_0$
component to be long-ranged. If the rotation is increased the loss of
invariance under the $C_2$ transform has a more significant effect on the
range of the $\sigma_0$ component and it vanishes more rapidly. This result
is shown in Fig. 2 of Ref. \onlinecite{bergeret} in which the Usadel
equation for this $S_s/F$ structure was solved, however, due to a spin
rotation of $\sigma_1$ the authors find the $\sigma_1$ component to be
long-ranged.

\section{Low energy density of states}

The full solution of the $\sigma $-model is obtained by considering
fluctuations about the saddle point solution. There are several different
types of fluctuations which are relevant to different cases. The low energy
C-mode fluctuations about the Usadel saddle-point solution are defined as
being diagonal in the advanced-retarded space and are therefore quantum
corrections to the Usadel solution. They have the further property that they
are independent of the order parameter and any magnetic field. The C-modes
dominate at energies below the Thouless energy $D/L^{2}$ where $L$ is the
length of the ferromagnet.~\cite{altland} We shall find the C-mode
fluctuations for an $S_{s}/F$ structure with $V=h(\sigma _{3}\cos \alpha
+\sigma _{2}\sin \alpha )$. We will then simplify our solution and provide a
qualitative description of the low energy density of states. We are
interested in seeing how the triplet component induced in the ferromagnet
affects the density of states. Our method closely follows that of Ref. %
\onlinecite{altland} where an $S_{s}/N$ structure was considered.

If the solution of the Usadel equation is $Q_{U}$ and we represent the
C-mode corrections by the matrix $T$ then the full solution of the
supermatrix is $Q=TQ_{U}T^{-1}$. One can show~\cite{altland} that at very
low energies the dominant C-mode is spatially constant, the so-called
zero-mode. In addition, $Q_U$ has a very slow spatial variation. The matrix $Q$ must satisfy the convergence symmetry and the
charge conjugation symmetry. The convergence symmetry is 
\begin{equation}
Q=\eta Q^{\dag }\eta ^{-1},\qquad \eta =E_{11}\tau _{3}\Lambda +E_{22}
\end{equation}
and the charge conjugation symmetry is 
\begin{equation}
Q=\tau Q^{T}\tau ^{-1},\qquad \tau =E_{22}i\rho _{2}+E_{11}\rho _{1}.
\end{equation}
We have defined 
\begin{equation}
E_{11}=\left( 
\begin{array}{cc}
1 & 0 \\ 
0 & 0
\end{array}
\right) _{{\rm bf}},\qquad E_{22}=\left( 
\begin{array}{cc}
0 & 0 \\ 
0 & 1
\end{array}
\right) _{{\rm bf}},
\end{equation}
where the subscript `${\rm bf}$' indicates boson-fermion space. Since $Q_{U}$
must also satisfy the above symmetries we may define the fluctuations as $%
T=e^{W}$ where $W$ must satisfy 
\begin{equation}
W^{\dag }=-\eta W\eta ^{-1},\qquad W^{T}=-\tau W\tau ^{-1}.
\end{equation}
The C-mode fluctuations must be insensitive to the superconducting order
parameter and magnetic fields so we require 
\begin{eqnarray}
\lbrack W,\sigma _{2}\tau _{1}\rho _{3}],\,[W,\sigma _{2}\tau _{2}]
&=&0,\qquad {\rm the\,\,order\,\,parameter\,\,commutes\,\,through}  \nonumber
\\
\left[ W,\tau _{3}\rho _{3}\right]  &=&0,\qquad {\rm the\,\,magnetic\,%
\,field\,\,commutes\,\,through.}
\end{eqnarray}
For a solution of $W$ we may use the zero-mode derived in Ref. %
\onlinecite{altland} but we must include some spin dependence 
\begin{eqnarray}
T &=&vua_{1}a_{2}a_{3}  \nonumber \\
a_{1} &=&\exp (i{\textstyle{\frac{1}{2}}}\theta _{1}E_{22}\tau _{1}\rho
_{1}\sigma _{1})  \nonumber \\
a_{2} &=&\exp (i{\textstyle{\frac{1}{2}}}\theta _{2}E_{22}\tau _{2}\rho
_{1}\sigma _{2})  \nonumber \\
a_{3} &=&\exp (i{\textstyle{\frac{1}{2}}}\theta _{3}E_{22}\tau _{1}\rho
_{2}\sigma _{3})  \nonumber \\
u &=&\exp (iyE_{22}\rho _{3})  \nonumber \\
v &=&\exp \left( 
\begin{array}{cc}
0 & \lambda -\mu \rho _{3} \\ 
\mu +\lambda \rho _{3} & 0
\end{array}
\right) _{{\rm bf}},  \label{eq:fluctuations}
\end{eqnarray}
where $y$ is some complex variable and $\lambda $ and $\mu $ are Grassmann
variables. The above solution is sufficiently general for our choice of $V$.
Terms which satisfy the symmetry requirements and are not included in $T$
are superfluous to our density of states calculation. We could have chosen,
for example, spin dependent fluctuations with the matrix structures $%
E_{22}\tau _{1}\rho _{2}\sigma _{1}$, $E_{22}\tau _{2}\rho _{2}\sigma _{2}$
and $E_{22}\tau _{1}\rho _{1}\sigma _{3}$ as they also satisfy the symmetry
requirements. However, they would add nothing extra to the final solution.
The extra terms will either vanish or make a contribution identical to the
one already obtained from $a_{1,2,3}$. One should note that the invariant
transform of the action of a singlet superconductor coupled to a ferromagnet
with $V=h(\sigma _{3}\cos \alpha +\sigma _{2}\sin \alpha )$ is $C_{1}$ and
that $T$ is also invariant under the $C_{1}$ transform. If we chose a
different exchange field, for example $V=h(\sigma _{3}\cos \alpha +\sigma
_{1}\sin \alpha )$ we should choose a different form of $T$. The above
choice of $a_{3}$ will not contribute to the action and should be replaced
with $\exp (i\frac{1}{2}\theta _{3}E_{22}\tau _{1}\rho _{1}\sigma _{3})$. In
this case the invariant transform of both the action and the fluctuations is 
$C_{2}$. Deriving a suitable form of $T$ can be quite tedious and the task
is considerably shortened if one chooses $T$ to have the same invariance
transform as the action under consideration. As stated above, this will not
give the most general form of $T$, but gives those parts which contribute
uniquely to the density of states.

%Our choice given in Eq. (\ref{eq:fluctuations}) is
%the simplest choice which minimises the number of vanishing %terms and
%ensures that the action remains energy and exchange field %dependent. Note
%that the action and the fluctuations chosen are all invariant %under the $C_{1}$ transform.

The solution of the Usadel saddle point equation is $Q_{U}=g_{0}$. One can
show that the part diagonal in particle-hole space which describes the
normal Green function is $g\tau _{3}$, i.e., $g=-g^{\dag }$. The
off-diagonals in particle hole space $f$ and $f^{\dag }$ describe the
anomalous Green function and may in general contain the terms $\tau _{1}$
and $\tau _{2}\rho _{3}$ multiplied by the spin components $\sigma _{0,1,3}$
and $\sigma _{2}\rho _{3}$. The spin components which actually appear in the
solution of $Q_{U}$ will of course be dependent on the spin structure of the
exchange field $V$. %of the form 
%$Q_{U}=q_{3}\tau
%_{3}+q_{22}\tau _{2}\sigma _{2}+q_{12}\tau _{1}\sigma _{2}\rho
%_{3}+q_{2}\tau _{2}\rho _{3}+q_{1}\tau _{1}$. The normal Green %functions
%are defined by $q_{3}$; $q_{22}$ and $q_{12}$ define the %anomalous singlet
%Green function and $q_{2}$ and $q_{1}$ define the long-ranged %anomalous triplet
%Green function. 
On substituting the general solution of $Q_{U}$ with the fluctuations $T$
into the action given in Eq. (\ref{eq:action}) with $V=h(\sigma _{3}\cos
\alpha +\sigma _{2}\sin \alpha )$ one finds that all the anomalous
components vanish. The singlet components vanish because they are
proportional to the order parameter which commutes with $T$ while the
triplet components give zero supertrace. One can show this is true even with
the most general form of $T$, which is why it is unnecessary to find the
most general form. %The anomalous singlet terms commute with $T$ (since the
%order parameter commutes through) so are negligible. After %substituting $%
%Q=TQ_{U}T^{-1}$ into the action given in Eq. (\ref{eq:action}) %with the
%singlet form of the order parameter one finds 
One finds the zero-mode action to be 
\begin{equation}
S=-2i\tilde{s}(\cos \theta _{1}\cos \theta _{2}\cos \theta
_{3}-1)+2ih_{1}\sin \theta _{1}\sin \theta _{2}\cos \theta _{3}-2ih_{2}\sin
\theta _{1}\sin \theta _{3}\cos \theta _{2}  \label{eq:C-action}
\end{equation}
where 
\begin{eqnarray}
\tilde{s} &=&\pi \epsilon \nu \int gdx  \nonumber \\
h_{1} &=&\pi h\nu \int g\cos \alpha dx  \nonumber \\
h_{2} &=&\pi h\nu \int g\sin \alpha dx.
\end{eqnarray}
%Note that the action only depends on the $\tau _{3}$ component %of the Usadel
%solution $q_{3}$. Also, 
Since the C-modes are diagonal in the advance-retarded space we need only
consider the retarded part so $Q$ has been reduced to a $16\times 16$
supermatrix and we may set $\delta =0$. The density of states with C-type zero-mode fluctuations is 
\begin{eqnarray}
\rho  &=&{\textstyle{\frac{\nu }{8}}}\,{\rm Re}\,\langle \,{\rm str}%
\,Q\sigma _{3}^{{\rm bf}}\tau _{3}\rangle   \nonumber \\
&=&2\nu \,{\rm Re}\,g\langle 1-{\textstyle{\frac{1}{2}}}(1-\cos \theta
_{1}\cos \theta _{2}\cos \theta _{3})+2\lambda \mu (1-\cos \theta _{1}\cos
\theta _{2}\cos \theta _{3})\rangle 
\end{eqnarray}
where the averaging is weighted by the action in equation (\ref{eq:C-action}%
) and we must perform a path integration over $Q$ (which means an integral
over the three $\theta $'s, $\lambda $, $\mu $ and $y$). This form of the
density of states and action is quite general and one would obtain something
similar for any exchange field of the form $V=h(\sigma _{i}\cos \alpha
+\sigma _{j}\sin \alpha )$.

As the path integration over all $Q$ is fairly complex we shall make an
approximation and assume that this solution will be at least qualitatively
similar to the full solution. We approximate by setting two of the $\theta$%
's to zero. The result is the same no matter which $\theta$ we choose to be
non-zero. In this simplified case our fluctuations are equivalent to those
in Ref. \onlinecite{altland} so we may assume the same Jacobian. The density
of states can be shown to be 
\begin{eqnarray}
\rho&=&2\nu\,{\rm Re}\,\left[g\left(1-{\textstyle{\frac{1}{2}}}
\int_0^{\pi}d\theta\sin\theta e^{2i\tilde{s}(\cos\theta_i-1)}\right)\right] 
\nonumber \\
&=&2\nu\,{\rm Re}\,\left[g\left(1-\frac{\sin 4\tilde{s}}{4\tilde{s}} -\frac{%
1-\cos 4\tilde{s}}{4i\tilde{s}}\right)\right].  \label{eq:dos}
\end{eqnarray}
This is the same formula as for an $S_s/N$ structure, but the Usadel
solution of an $S_s/F$ is not the same as for an $S_s/N$ so the value of $%
\tilde{s}$ will be different. An exact solution of the Usadel equation for
an $S_s/F$ structure with a non-homogeneous exchange field does not exist.

We now calculate an approximate solution of the density of states in the
region where the ferromagnet is homogeneous $w<x<L$. To find a solution for
the density of states we use the approximate solution for the Usadel
equation derived in Ref. \onlinecite{bergeret} which is valid when the
tunneling resistivity $R_b$ from the superconductor to the ferromagnet is
large (it is also valid near the phase transition but this requires $%
\Delta\ll\epsilon$ which does not satisfy our small energy requirement). In
this limit the bulk solution may be taken in the superconductor $g=\epsilon/%
\sqrt{\epsilon^2-|\Delta|^2}$ which is vanishingly small. In the ferromagnet
limit we can set $g\sim 1$ and then the anomalous Green functions $f$ can
be calculated from the linearised Usadel equation (\ref{eq:usadel}). We may
obtain a solution for $g$ from the identity $g_0^2=1$ which implies $g=\sqrt{%
1-f^{\dag}f} \sim 1-\frac{1}{2} f^{\dag}f$. In Ref. \onlinecite{bergeret} it
is shown that $f$ contains both a singlet component and a triplet component.
If the ferromagnetic exchange field is large compared to the energy then the
singlet part is much smaller than the triplet in the region $w<x<L$ so may
be neglected. The coefficient of the triplet component is derived in Ref. %
\onlinecite{bergeret} although some care must be taken as one must perform
two rotations to make it compatible with the matrix structures used here.
The result is, when taking just the triplet component, $f^{\dag}f\sim -C^2$
where 
\begin{equation}
C^{R,A}=\mp iAB(0)\sinh[\kappa_{\epsilon}(L-x)] [\kappa_{\epsilon}\cosh%
\Theta_{\epsilon}\cosh\Theta_3 +\kappa_3\sinh\Theta_{\epsilon}\sinh\Theta_3%
]^{-1}
\end{equation}
for $w<x<L$ and where $B(0)=(\rho\xi_h/2 R_b)f_s$, $f_s=\Delta/\sqrt{%
\epsilon^2-\Delta^2}$, $\Theta_{\epsilon}=\kappa_{\epsilon}L$, $%
\Theta_3=\kappa_3 L$, $\kappa_3=\sqrt{A^2+\kappa_{\epsilon}^2}$. To
calculate the action we require 
\begin{equation}
\tilde{s}=\pi\epsilon\nu\left(\int_{-\infty}^0gdx+\int_0^wgdx
+\int_w^Lgdx\right).
\end{equation}
In the small energy limit $g$ is very small in the superconductor so we will
neglect the integral over negative $x$. If we assume that $w$ is small then
we may also neglect the second integral. So now $\tilde{s}$ just depends on
the value of $g$ in the homogeneous part of the ferromagnet which we have
found to be 
\begin{equation}
g\sim 1-{\textstyle{\frac{1}{2}}} A^2B(0)^2\sinh^2
[\kappa_{\epsilon}(L-x)]\left( \kappa_{\epsilon}\cosh
L\kappa_{\epsilon}\cosh w \sqrt{A^2+\kappa_{\epsilon}^2}+\sqrt{%
A^2+\kappa_{\epsilon}^2} \sinh L\kappa_{\epsilon}\sinh w\sqrt{%
A^2+\kappa_{\epsilon}^2}\right)^{-2}
\end{equation}
and therefore 
\begin{eqnarray}
\tilde{s}&=&\pi\epsilon\nu(L-w)\left(1+{\textstyle{\frac{1}{4}}} A^2B(0)^2
\left(1-{\textstyle{\frac{1}{2}}}(L-w)^{-1}\kappa_{\epsilon}^{-1} \sinh[%
2\kappa_{\epsilon}(L-w)] \right)\right.  \nonumber \\
&&\qquad\times\left.\left(\kappa_{\epsilon}\cosh L\kappa_{\epsilon}\cosh w 
\sqrt{A^2+\kappa_{\epsilon}^2}+\sqrt{A^2+\kappa_{\epsilon}^2} \sinh
L\kappa_{\epsilon}\sinh w\sqrt{A^2+\kappa_{\epsilon}^2} \right)^{-2}\right).
\end{eqnarray}
Substituting $\tilde{s}$ into equation (\ref{eq:dos}) gives the low energy
density of states within the homogeneous part of the ferromagnet ($x>w$). We
find that the density of states vanishes at $\epsilon=0$ indicating the
presence of a micro-gap and the low energy behaviour is quadratic in $%
\epsilon$. This is similar to what has been found in an $S_s/N$ structure.~\cite{altland,frahm} Note that in an $S_s/F$ structure where there is no
long-ranged anomalous Green function in the ferromagnet one would not
expect a mirco-gap in this high tunnelling resistivity limit. In this case $%
\tilde{s}\sim \pi\epsilon\nu L$ and as $L\rightarrow\infty$ the density of
states is simply $\rho=2\nu$, i.e., the bulk normal-metal density of states.

An equivalent calculation for an $S_t/F$ structure is much simpler. The
C-mode fluctuations are defined to commute with the order parameter so in
the case of a triplet superconductor these fluctuations must be independent
of spin. Therefore an $S_t/F$ is similar to an $S/N$ and one can show that
equation (\ref{eq:dos}), which is exact for $S/N$ but an approximation for $%
S_s/F$, is exact for $S_t/F$. One can solve the linear Usadel equation in
the ferromagnet to show that the form of the low energy density of states is
the same as in the normal metal of a $S/N$ structure, displaying a micro-gap
as the energy vanishes.

\section{Conclusion}

We have considered an unusual type of triplet Cooper pairing which is
defined by an order parameter which is even in the momentum (or position)
and odd in the frequency (or time). It was found that, for the most part, a
superconductor with odd triplet Cooper pairs is much like the standard
singlet superconductor (even in position and time). In the bulk these
superconductors would appear to be much the same, and also when coupled to a
normal metal. The main difference between the two superconductors is their
spin structure. Another difference is the energy dependence of the order
parameter though, in many cases, this is not important.

If we consider a situation where the spin is unimportant we may obtain
equations for $S_{t}$ from equations for $S_{s}$ by simply replacing the
order parameter $\Delta $ with $\,{\rm sgn}(\epsilon )\Delta $. However, in
density of states calculations, for example, this change of sign is
irrelevant. Where we do observe a difference between the $S_{t}$ and the $%
S_{s}$ is in cases where the spin is important. When an $S_{s}$ is coupled
to an inhomogeneous ferromagnet it is possible to induce a long-ranged
triplet anomalous Green function component as well as a short-ranged
singlet component in the ferromagnet. However, when an $S_{t}$ is coupled to
any type of ferromagnet a long-ranged triplet component always exists in the
ferromagnet. One can determine which anomalous components will dominate the
ferromagnet by considering the transformational invariances of the $\sigma $%
-model. We considered the low-energy fluctuations about the Usadel solution
of an $S_{s}/F$ structure with a non-homogeneous exchange field in order to
see if the long-range triplet has a significant effect. We found that an $%
S_{s}/F$ structure which induces a long-range anomalous component in $F$
will have a density of states which vanishes quadratically at small
energies. This micro-gap in the density of states is also expected in $%
S_{s}/N$  and $S_t/F$ structures but not in $S_{s}/F$ structures which do not have any
long-ranged anomalous components.

\begin{acknowledgments}
We are grateful to F. S. Bergeret and A. F. Volkov for useful discussions.
\end{acknowledgments}

%\begin{references}

%\end{references}

\end{document}